\documentclass[11pt]{article}
\usepackage{graphicx,xspace,amsmath}
\usepackage{xcolor}
\usepackage{lineno}
\usepackage[authoryear]{natbib}
\newcommand{\ie}{\emph{i.e.}\xspace}

\newcommand{\etc}{{\em etc.\/}\xspace}

\newtheorem{rem*}{Remark}

\newcommand{\figref}[1]{\mbox{Figure~\ref{fig:#1}}}
\newcommand{\tabref}[1]{\mbox{Table~\ref{tab:#1}}}

\newcommand{\cd}{\,|\,}
\renewcommand{\H}{\mathcal{H}}
\newcommand{\Hp}{\H_{\text{p}}}

\newcommand{\LR}{\text{LR}}

\newcommand{\Hugin}{{\tt Hugin}}

\begin{document}
\title{{\sc The Epistemic Value of Novel Predictive Success in Scientific and Criminal Investigations: A Bayesian Explanation}}
\author{
Julia Mortera\thanks { \hspace*{5mm} Email: {\tt julia.mortera@uniroma3.it}}\\
University of Bristol, UK. \\
\and William C. Thompson\thanks {
 \hspace*{5mm} Email: {\tt william.thompson@uci.edu}.}\\
University of California Irvine, USA.\\
 }
\date{\today}
\maketitle
\begin{abstract}
Because there are similarities between the evaluation of alternative stories in criminal trials and the evaluation of scientific theories, scholars have looked to literature in epistemology and the philosophy of science for insights on the evaluation of evidence in criminal trials.  The philosophical literature is divided, however, on a key point -- the epistemic value of novel predictive success.  This article uses a Bayesian network analysis to explore, in the context of a criminal case, the circumstances in which “new evidence” discovered after a theory is propounded, can provide stronger (or weaker) support for the theory than “old evidence” that was accommodated by the theory.  It argues that insights from analysis of the strength of “new” and “old” evidence in the criminal case can be applied more generally when assessing the relative merits of prediction and accommodation in scientific theory development, and are thus helpful in addressing the longstanding philosophic controversy over this issue.
\end{abstract}
\noindent {\small {\em Some key words:} Bayesian networks, criminal investigation, confirmation theory, 
heterogenous population, likelihood ratio, novel facts, strength of evidence.}

\section{Introduction}

In adversarial legal systems, the burden of proof in criminal cases rests firmly on the prosecution.  To meet this burden, the prosecutor must tell a convincing story in which the defendant bears responsibility for a crime; it must support this story with evidence.  The defendant is not required to give an account of the evidence; the defence can simply challenge the credibility of the prosecutor’s account.  Quite often, however, the defence will offer an alternative story in which the defendant either did not commit a crime or committed a less serious offense than the prosecutor has charged.  The trier-of-fact must then consider whether the prosecutor’s story is sufficiently convincing to meet the burden of proof, or whether the defence story, or some other account of the evidence, is sufficiently plausible to raise doubts about the defendant’s guilt.  

Philosophers have noted similarities between the evaluation of alternative stories in criminal trials and the evaluation of scientific theories.   There is a deep literature in epistemology and the philosophy of science on how to evaluate scientific theories and determine when they are trustworthy.  Scholars have looked to this literature for insights on the evaluation of alternative theories offered by prosecution and defence in criminal cases.  An interesting example is Anne Ruth Mackor's  discussion of the epistemic value of novel predictive success as a criterion for assessing such theories \citep{Mackor2017}.

Mackor noted that the ability of a scientific theory to generate novel (i.e., previously untested) predictions that turn out to be true has long been viewed as an important indicator that the theory is correct. This viewpoint, known as predictivism, was advocated with particular fervor by  \citet{lakatos1978science} and his followers, who argued that a program of scientific research must generate novel predictions to be considered “progressive.” Applying Lakatos' perspecitve to criminal law, Mackor argued that theories introduced by a party gain in credibility if they predict new facts that are later found to be true.  This is particularly true if the predictions are “risky,” which means that they are unlikely if the underlying theory is false.  These predictions pass what philosophers call the “no miracles argument test”: if it would be miraculous to find that a prediction happened to be true if the underlying theory was false, then the successful prediction strongly supports the theory’s truth.    

Another argument for predictivism is that it provides insurance against certain questionable research practices (QRPs) that may undermine the epistemic value of “old”  evidence \citep{syrjanen2023novel}, such as overfitting a theory to a limited set of evidence \citep{hitchcock2004prediction}, “hypothesis hunting,” (Mayo, 1996, pp. 294-318) or “fudging” a theory in an effort to accommodate a particular datum \citep{lipton2005testing}. Mackor argued that an analogous process can occur in criminal cases.  Indeed, this problem may be particularly acute in criminal trials because parties are often motivated to invent convenient stories.  A guilty defendant, for example, may be highly motivated to generate a false story that explains as much of the incriminating evidence as possible in a non-incriminating manner.  If those stories are untrue, of course, they are unlikely to successfully predict new evidence.  Consequently, there is special epistemic value to finding new evidence consistent with such a theory.  

While Mackor's analysis seems sound and insightful, there are difficulties with using philosophical analysis as a source of guidance for evaluating criminal cases.  One problem is that philosophers often disagree.  For example, in a comprehensive review article \citet{douglas2013state} concluded that there is “considerable disagreement about the epistemic value of novel predictive success…”  Some scholars, known as accommodationists, contend that the evidentiary support provided by scientific evidence depends solely on its logical relationship to the theory being evaluated, and not on whether the evidence was collected before or after the theory was proposed.  Douglas and Magnus note a “long-standing debate” between philosophers who “subscribed to the idea that predicting data was ceteris paribus better than accommodating it” (e.g., Leibniz, Whewell, Duhem) and those who disagreed (e.g., Mill and Keynes).   \citet{hitchcock2004prediction} offer the following account of the respective positions: 
\begin{quote}
	
	This issue has engaged philosophers of science for more than a century and a half. \citet{whewell1840aphorisms} emphasized the importance of predictive novelty in his philosophy of inductive science, while Mill (1843) claimed that no serious scientific mind could grant more than a psychological distinction between prediction and accommodation.  The issue drives a wedge between what Musgrave (1974) calls logical and historical theories of confirmation.  According to the former, the extent to which [data confirm theory] is purely a function of the logical and mathematical relationship [between theory and data].  Logical theories of confirmation include Hempel’s (1945) theory of instance confirmation, Glymour’s (1980) bootstrap theory of confirmation, and the likelihood approach of Edwards (1972) and Royall (1997).  By contrast, historical theories of confirmation assert that the time at which a theory was propounded, and even the thought process that went into its construction, can affect the theory’s epistemic status.  Popper’s falsificationist methodology (Popper, 1959) fits this model. (\citet{hitchcock2004prediction}, p. 2)
\end{quote}

In recent years, philosophical debate “has shifted away from Lakatosian accounts of theory change and assessment of research programs to whether formal theories of confirmation (especially Bayesian) could accommodate predictivist intuitions…”  \citep{douglas2013state}.  Here also philosophers have been divided, with disagreement on at least two issues—whether Bayesian analysis can explain predictivist intuitions and whether Bayesian analysis is valid.  According to \citet{brush1994dynamics}, ‘‘philosophers have defended all four possible positions: Bayesian analysis is (i) valid because it favors novel prediction, (ii) valid because it does not favor novel predictions, (iii) invalid because it favors novel predictions, and (iv) invalid because it does not favor novel predictions’’ (Brush, 1994, p. 134, quoted in Douglas and Magnus, 2013). Needless to say, such disagreement poses a challenge for those who look to philosophy for guidance on the epistemic value of novel predictive success, whether in scientific or criminal investigations.  

Among predictivists, there is also disagreement (and a degree of vagueness) regarding the reasons for favoring novel evidence.  Douglas and Magnus note an emerging viewpoint that rejects the idea that novel predictions have “intrinsic epistemic virtue” in favor of the idea that novelty is simply a proxy or indicator of other factors that may provide assurance that a theory is correct.  In other words, novel evidence is good because it is often (although not always) associated with other factors that make the underlying theory more trustworthy.  They call this position Plural Instrumental Predictivism (PIP), although others have called it “weak” or “local” predictivism \citep{hitchcock2004prediction,barnes2022prediction}. Whatever the merits of this position, it has not yet provided a full explanation of the “other factors” that may assure the trustworthiness of underlying scientific theories nor a method for determining when and whether they apply.   

\citet{syrjanen2023novel} offers what may be the most complete analysis of circumstances in which novel prediction can be superior to accommodation.  He argues, however, that the reverse may also occur because  QRPs may undermine the epistemic value of data collected after a theory as well as data collected before the theory was developed.  Indeed, according to Syrj{\"a}nen, “novel prediction and accommodation appear roughly on a par, or accommodation is even superior in the current context.” (p. 181).  He argues, and we agree, that evaluation of a scientific theory requires a close examination in each instance of the logical connections between the theory and the evidence, and that the distinction between novel prediction and accommodation is not always relevant to the strength of those connections.   \citet{jellema2021values} offers a similar analysis concerning the value of prediction in criminal cases.

This article is designed to cast new light on these complex and vexing questions using Bayesian networks.  We will argue that Bayesian networks are helpful for understanding the epistemic value of novel predictive success--\ie, why, when, and how much our belief in a theory should be influenced by the ability of the theory to predict an unexpected finding. We contend that past skepticism about the value of Bayesian analysis for addressing these issues is unwarranted.  For example, the widely-discussed claim that Bayesian analysis gives no weight to “old evidence” \citep{glymour1980discussion,barnes2022prediction}
is simply incorrect—a point we will demonstrate by example (while offering a more analytic discussion in Appendix 1).  We will argue that properly designed Bayesian networks produce results that are largely consistent with philosophical intuitions about the value of novel evidence, while at the same time explicating the circumstances in which such evidence may deserve more or less weight.   

To illustrate the Bayesian network approach, we will analyze a criminal case involving “novel” evidence.  Our case is hypothetical, although it closely follows the facts of a California case in which one of the authors (WT) was retained by defence counsel to assist in evaluating DNA evidence against the defendant.  After using this case to explain circumstances in which novel predictive success is (and is not) important in a criminal investigation, we apply our analysis to theory testing more generally.

\section{ An Illustrative Case Study   }
\label{sec:motivating}

The defendant was arrested and charged with murder after a police officer saw him running from a public park where a man had just been murdered.  The victim, known to police as a drug dealer, had been stabbed to death.  

DNA tests found that blood on the defendant’s clothing matched the DNA profile of the victim.  There were no witnesses to the crime, which occurred at night in a dark area of the park.  Police interviewed a local resident who reported seeing the drug dealer and the defendant, but no one else, near the park shortly before the murder.  The defendant refused to answer any questions put to him by the police.

The prosecution proceeded on the theory that the defendant had killed the drug dealer in a dispute over a drug transaction.  The defendant’s lawyer offered an alternative theory: that defendant had gone to the park to buy drugs and had been a witness when a third man appeared and stabbed the drug-dealer. According to the defence hypothesis, the defendant had blood on his clothing because the victim had stumbled against him before falling to the ground; and the defendant ran from the scene due to fear that the killer would attack him as well.  

Thereafter, two new pieces of evidence emerged.  We will argue that the strength of the case against the defendant varies greatly depending on the order in which this new evidence was discovered and became known to relevant actors.  Please consider at this point how and whether the order of discovery affects your intuitive evaluation of the defendant’s guilt.

One new item of evidence was the defendant's description of the killer.  He said that the killer was “a huge Samoan guy.”  In the city where the crime occurred (Long Beach, California) Pacific Islanders constitute approximately one percent of the population.  Samoan men, who are part of that sub-population, tend to be large and muscular.

The other new item of evidence came from a consultant who reviewed the DNA evidence at the request of the defence lawyer.  The consultant said that the crime lab had performed DNA tests on blood stains found at the crime scene.  The DNA profile on most of these items matched the victim, but one of the items had a different DNA profile that matched neither the victim nor the defendant.  The genetic alleles that constitute a DNA profile vary in their frequency of occurrence in various human populations.  The consultant used a worldwide database of allele frequencies to calculate the expected frequency of the unknown DNA profile in various human populations, including Samoans.  He reported that the unknown profile would be extremely rare in the Hispanic, Caucasian or Afro-American populations of the United States, and most other human populations, but would be much more common among Samoans.   {{Table \ref{tab:freq} in Appendix 2 gives the frequencies of the alleles in the DNA profile in the different populations.  }}

How does this new evidence affect our evaluation of the theory that the defendant is the murderer?  We will argue that both our intuitions and a formal Bayesian analysis suggest that the answer depends on the order in which the new items of evidence became known—and specifically, whether the defendant knew about the defence consultant’s evidence before telling his story about the huge Samoan.  If he did not know about the consultant’s evidence, then the defendant was advancing a theory of the case that included a rather improbable element—that the killer was Samoan.  The consultant’s evidence, coming later, can be seen as independent confirmation (or at least support) of this improbable element: it unexpectedly supported what might be seen as a “risky prediction”.  Our intuitions suggest that the evidence, coming in this order, makes it more probable that the defendant's story about the huge Samoan was truthful, and hence less likely that he is guilty.

The defendant's story is less convincing, however, if he knew about the consultant’s report beforehand.  In that case, defendant may simply have made up a convenient story—advancing a theory designed to fit the evidence in a manner consistent with his innocence.  The key element of his story—that the killer was Samoan—is no longer a “risky prediction” because he already knew that the DNA evidence supports that theory.  In this case, we think most people will agree, the defendant’s story is less believable and hence he is more likely to be guilty.      

In the next section we will present a Bayesian network model that uses inferential logic to explain how the strength  of the case against the defendant {{varies based on the unfolding of the different pieces of evidence.}}  We will then offer a broader discussion of the circumstances in which the epistemic support for a theory will and will not depend on the way in which the theory was developed.

\section{Introduction to Bayesian Networks}
\label{sec:BN}

Bayesian networks (BNs) have been widely used to model legal and forensic reasoning 
see \citet{dmpb:sjs,taroni06,dawid2020bayesian} and references therein. 
A Bayesian Network \citep{cowell99} is a graphical model describing the various items
of evidence and hypotheses, and the probabilistic relationships
between them.  Such a representation  clearly displays the relevance of
the evidence to hypotheses of interest, and supports efficient computational 
algorithms to compute the impact of the evidence.    The hypotheses, items of evidence, and other variables are represented by nodes in the network.  The probabilistic relationships among the nodes are indicated by arrows.  The arrows go from  “parent” node(s) to a “child” node(s) and indicate the probabilistic dependence of each child on it's parents. Each node of a BN is associated with a  table describing the conditional probability distribution over its states, conditional on the state(s) of its
parent node(s). A BN supports inference about the
impact of observed evidence on hypotheses of interest. 

Once the qualitative structure has been built, the network is then populated with  quantitative information about those probabilistic dependencies. Once the BN is set
up in a suitable software environment, one enters (or instantiates) the available evidence on
the observed variables, and the system will compute the resulting 
posterior probability of hypotheses of interest given all the evidence. The interpretation is that each variable
depends on those  variables which feed an arrow into it (its ``parents''),
but not otherwise on earlier variables.

A useful extension of the idea of a Bayesian Network is the Object-Oriented
Bayesian Network (OOBN). This organises the nodes into a hierarachy of sub-networks, which can greatly simplify specification and interpretation.

Bayesian networks are not merely representations of the elements of logical inference, they are also tools that can be used for making inferences.  Software has been developed that allows users to explore the inferential connections between the variables in the network.  For example, one can enter evidence on particular variables  in order to see how this affects other variables in the network. 

 We will be using a BN to see how various items of evidence affect the posterior probability that a defendant is guilty.  The software shows how a rational observer, that agrees with the BN model representing the case, should update her beliefs in light of various items of evidence.  We will use a BN model to evaluate our illustrative case and show how, as we learn   each item of evidence, this affects  the strength  for incriminating or exonerating  the defendant.

\section{Bayesian networks representing the motivating example about novel facts}
\label{sec:BNnovel}

\begin{figure}[htbp]
	\begin{center}
		\caption{Pictorial representation of the Bayesian Network integrating the DNA findings to the case narrative.} 
		\label{fig:global}
		\includegraphics[width=.90\textwidth]{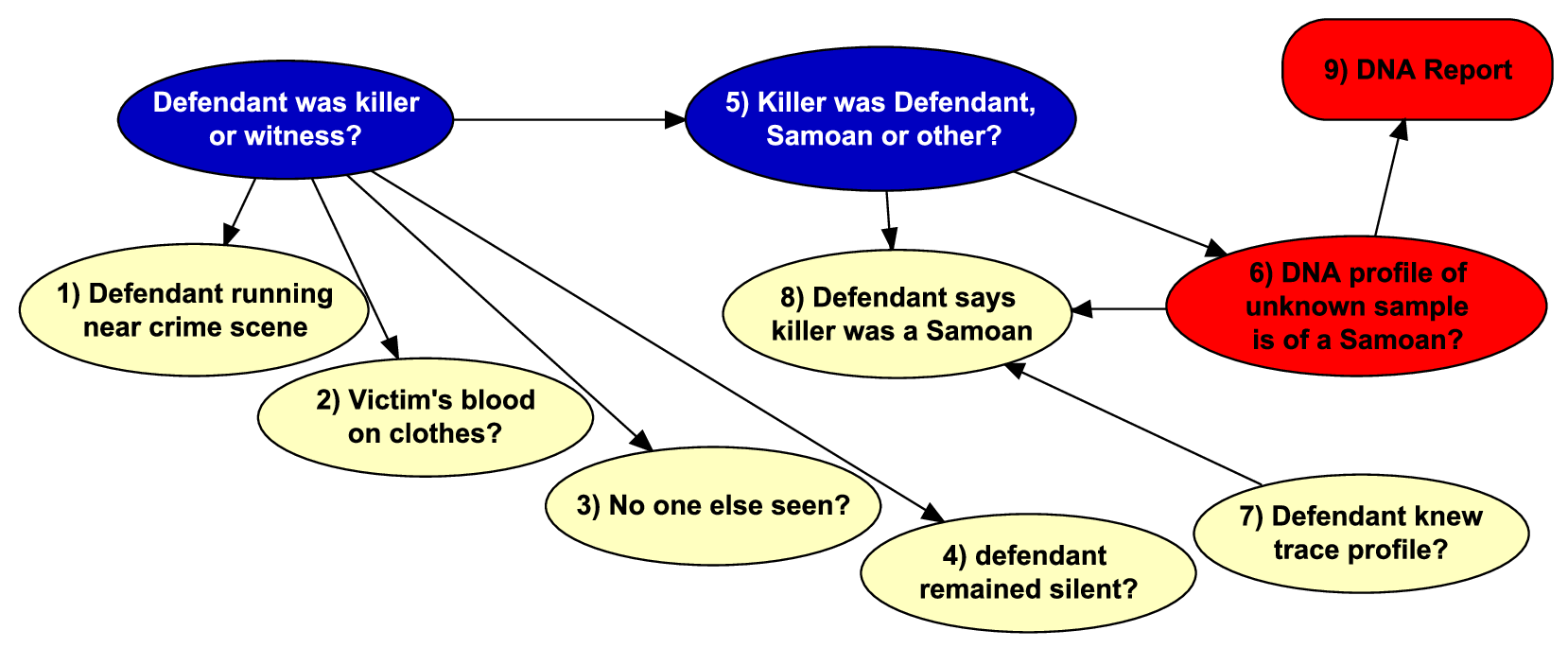}
	\end{center}
\end{figure}

The network in \figref{global} gives an overall pictorial representation of the problem. It integrates the report on the DNA findings on the trace found at the scene of the crime with a  network that represents the narrative of the case. The details of the DNA analysis, shown in \figref{global} as a red oval node  \texttt{9) DNA report} will be given in Section \ref{sec:DNA} and detailed in Appendix 2.
The   upper left node in \figref{global} represents two key alternative hypotheses about the case -- the prosecution 
hypothesis that the defendant was the killer and the defence hypothesis that the defendant was merely a witness. We assign equal prior probabilities to the two alternative hypotheses.  This node has four “children” that represent different items of evidence, \ie that the defendant was:
\begin{description}
	\item[1)] running near the scene of the crime; 
	\item[2)]  with the victim’s blood on his clothing; 
	\item[3)] that a witness saw no one other than the defendant and victim in the area; 
	\item[4)] that the defendant refused to respond to police questions.  
\end{description}
 
 We view these items as conditionally independent given the hypothesis, which means  each item depends directly only on which hypothesis is true, and not on the other items of evidence.  Thus, a single arrow links the hypothesis to each item  and there are no arrows between the four items. However, evidence in any item will affect the other items through  the posterior probabilities of the hypotheses parent node. 

 Table \ref{tab:CPT1} shows the conditional probabilities we assigned to the four evidence items.  These represent our subjective judgment of the probability that each item of evidence would be true, conditioned on the two hypotheses.  For example, we thought there was a 90\% chance that the defendant would have the victim’s blood on his clothing if he was the killer, and an 80\% chance if he was, as he claims, a witness.  We thought it was about 10 times more likely that no one (other than defendant and victim) would be seen if the defendant was the killer than if the killer was a third person, so we assigned probabilities of 0.9 and 0.1 respectively.  Readers are of course free to disagree with our subjective assignments and to try out different probabilities.  The goal of this exercise is to explore the probative value of the evidence in this case, given the conditional probabilities.  In most instances the exact conditional probabilities assigned make little difference to the issues we are exploring here.

\begin{table}[htbp]
\centering
\caption{Conditional probability table of the left four nodes in the Bayesian network shown in \figref{global}.}

\begin{tabular}{|c|cc|}
	\hline
	& \multicolumn{2}{c|}{If the defendant was:}           \\
	\hline
	Conditional probability that: & the killer &  a witness \\
	\hline
	Defendant was running near scene &       0.99 &        0.8 \\
	\hline
	Victim's blood on his clothes &        0.9 &        0.8 \\
	\hline
	No one else seen &        0.9 &        0.1 \\
	\hline
	Defendant remained silent &        0.9 &        0.3 \\
	\hline
\end{tabular}  
 \label{tab:CPT1}%
\end{table}%

\begin{table}[htbp]
	\centering
	\caption{Conditional probability table of the top right blue node of the Bayesian network shown in \figref{global}.}
	\begin{tabular}{|r|rr|}
		\hline
		& \multicolumn{2}{c|}{If the defendant was:}           \\
		{ Murderer} & { killer} & {witness} \\
		\hline
		defendant &    { 1} &    { 0} \\
		other &    { 0} & { 0.99} \\
		samoan &    { 0} & { 0.01} \\
		\hline
	\end{tabular}  
	\label{tab:prior}%
\end{table}%
\begin{table}[htbp]
	\centering
	\caption{Conditional probability table for node “defendant says killer was a Samoan”  given its parents in the Bayesian network shown in \figref{global}.}
\small{\begin{tabular}{|l|rr|rr|rr|rr|rr|rr|}
\hline
	&  \multicolumn{12}{c|}{Killer was} \\
	        &   \multicolumn{4}{c|}{Defendant } &             \multicolumn{4}{c|}{Other}    & \multicolumn{4}{c|}{A Samoan }          \\
	        \hline
	        	DNA Report supports            &        &            &       &            &        &            &       &            &       &            &      &            \\
	 Samoan Theory   &                 \multicolumn{2}{c|}{True (T)}  &                \multicolumn{2}{c|}{False (F)}   &            \multicolumn{2}{c|}{True (T)} &            \multicolumn{2}{c|}{False (F)} &            \multicolumn{2}{c|}{True (T)} &            \multicolumn{2}{c|}{False (F)}             \\
	 \hline
Defendant knew about          &        &            &       &            &        &            &       &            &       &            &      &            \\
	 DNA Report          &       T &      F &       T &      F &       T &      F &       T &      F &       T &      F &       T &      F \\
	 \hline
	 Probability defendant says          &          1 &        0.05 &       0 &        0.05 &        0.5 &      0 &        0 &          0 &          1 &          1 &        0.99 &        1 \\
killer was a Samoan	            &        &            &       &            &        &            &       &            &       &            &      &            \\
	\hline
\end{tabular}  }
 \label{tab:CPT2}%
\end{table}%

The detailed documentation of the BN that includes all the conditional probability tables, is available  at \begin{verbatim}https://www.dropbox.com/scl/fo/zvxth8gnfnzxbep7z7kyt/AOh-AHx-q6v5rSi2kVpe3jY?rlkey
=t9q9ndsjr7nebpp3nlltjc8po&dl=0
\end{verbatim}

The node at top center of the network subdivides the underlying hypotheses into three mutually exclusive and exhaustive propositions: the killer was either (i) the defendant; (ii) a Samoan; or (iii) someone other than the defendant or a Samoan. This node has conditional  probabilities given in  Table \ref{tab:prior} which reflect the fact that if the defendant was not the killer it is quite unlikely a priori, 1\%,  that it is a Samoan, given that this is roughly the frequency of this ethnicity in the relative population\footnote{\texttt{https://en.wikipedia.org/wiki/Long\_Beach,\_California}}.   This node has two children that represent items of evidence.  One child is a node representing the conditional probabilities based  on the unknown  profile at the crime scene that this profile is from a Samoan (or the alternative, that it is not from a Samoan). The DNA analysis of this unknown profile indicated by the Object-Oriented Bayesian Network (OOBN) node \texttt{DNA Report} is given in Section \ref{sec:DNA} and detailed in Appendix 2.   For present purposes, it is sufficient to know that the DNA profile of the unknown profile found at the crime scene was approximately 300 times more likely if it came from a Samoan than if it came from another ethnical group. Hence  the DNA analysis strongly supported the theory that the unknown DNA was that of a Samoan.   

The second child of the top-center node represents the defendant's statement that the killer was a Samoan.  This is a complex node because it has three parents, which indicates that the conditional probability of the defendant making this statement depends on these three variables: 
\begin{description}
	\item[5)] whether the defendant was the killer; 
	\item[6)]  whether the unknown DNA profile was that of a Samoan;
	\item[7)]  whether the defendant knew the results of the DNA test on the unknown profile. 
\end{description}
  Table \ref{tab:CPT2} shows the conditional probabilities we assigned to this node.  These probabilities reflect our judgment about the likelihood of the defendant claiming the killer was a Samoan under 12 distinct circumstances defined by all possible states of the three parent nodes.  
It is important to notice that the two children of the top-center node and nodes 6) and 7)  are not conditionally independent.  The arrow that connects these two nodes indicates that the probability of the defendant claiming the killer was a Samoan may depend partly on whether the unknown profile was from a Samoan, which might happen if the defendant became aware of the DNA report before offering his story about the Samoan.

In Table \ref{tab:CPT2}  we set the conditional probability of the defendant claiming the killer was a Samoan as certain, if  the DNA report confirms this, when he knows about the report   (columns 11) . 
 We think it is unlikely that the defendant would report that the killer was a Samoan if he saw someone other than a Samoan commit the murder.  So, we put the probability at zero for all of those instances (columns 7--9).  

If the defendant actually committed the murder, but the DNA report supported the theory that the unknown profile was of a Samoan, and the defendant knew about that, then we think it basically certain that he would falsely claim the killer was a Samoan in an effort to save himself from being convicted of murder.  On the other hand, we think it  less likely that he would make this claim if the DNA report did not support the Samoan theory.   As Samoans constitute only about 1\% of the local population, we considered assigning a probability of 1\% in this instance, but we thought stereotypes about Samoan men being big and tough might increase the chances that the defendant would choose a Samoan protagonist for a fanciful, lying account of what happened.  Hence, we put those probabilities at 5\%. 

\begin{figure}[htbp]
	\begin{center}
		\caption{Bayesian network prior to any evidence.} 
		\label{fig:naive}
		\includegraphics[width=1.0\textwidth]{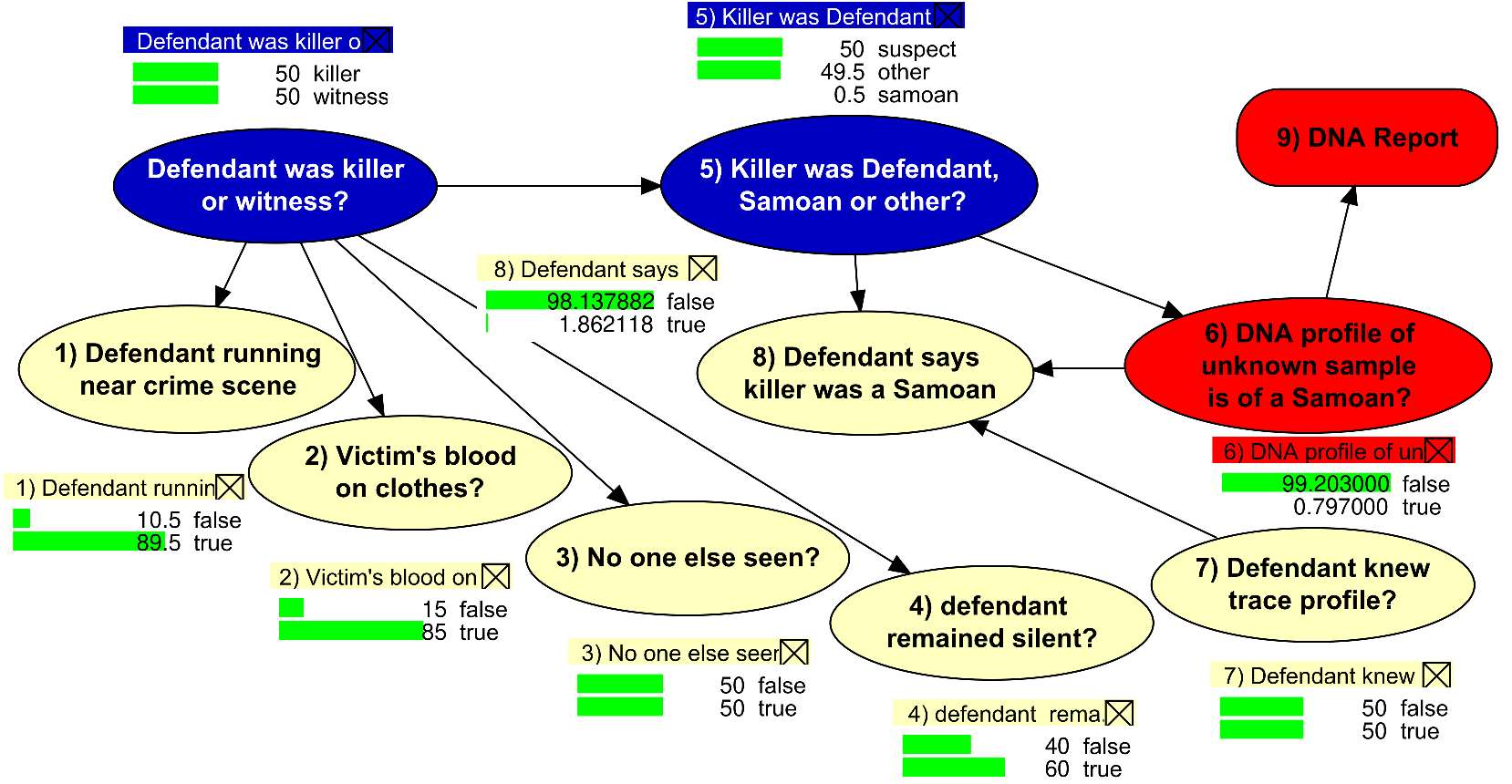}
	\end{center}
\end{figure}

\begin{figure}[htbp]
	\begin{center}
		\caption{Updated probabilities conditioned on evidence of the defendant running away from the crime scene.} 
		\label{fig:run}
		\includegraphics[width=.90\textwidth]{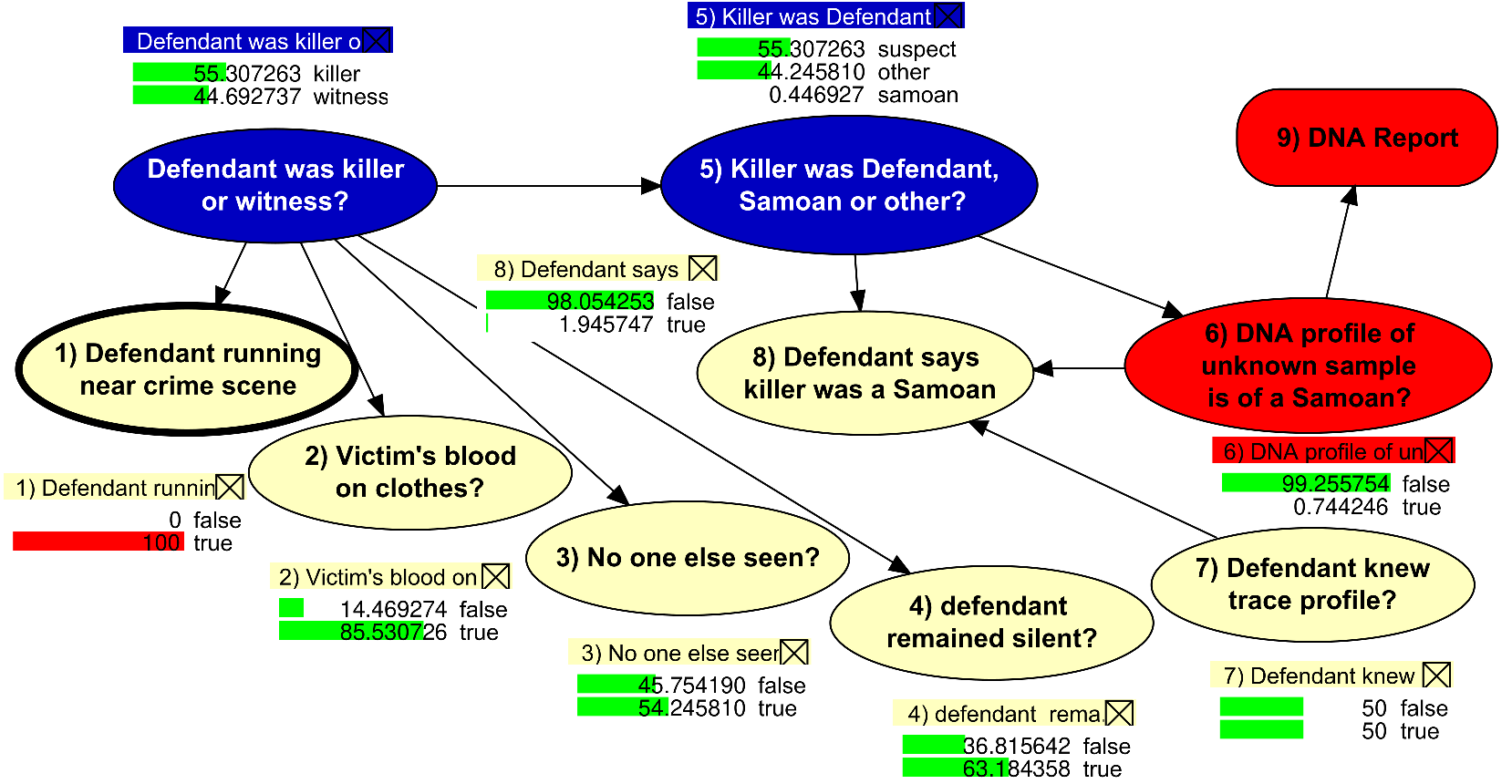}
	\end{center}
\end{figure}

\begin{figure}[htbp]
	\begin{center}
		\caption{Updated probabilities conditioned on evidence of the defendant running away from the crime scene and having victim's blood on his clothes.} 
		\label{fig:blood}
		\includegraphics[width=.90\textwidth]{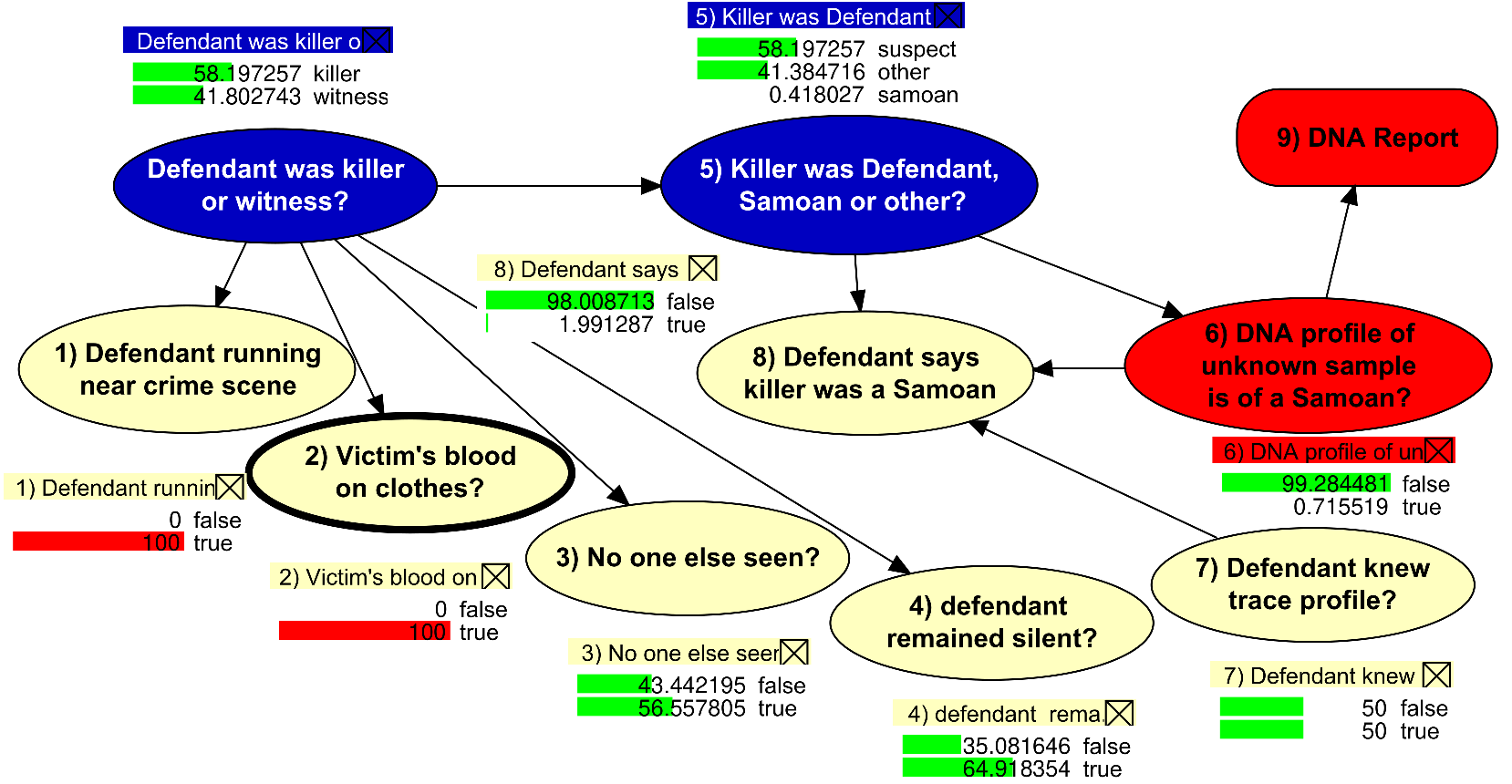}
	\end{center}
\end{figure}

\begin{figure}[htbp]
	\begin{center}
		\caption{Updated probabilities conditioned on evidence of the defendant running away from the crime scene and having victim's blood on his clothes and not being anyone else seen at the crime scene.} 
		\label{fig:noone}
		\includegraphics[width=.90\textwidth]{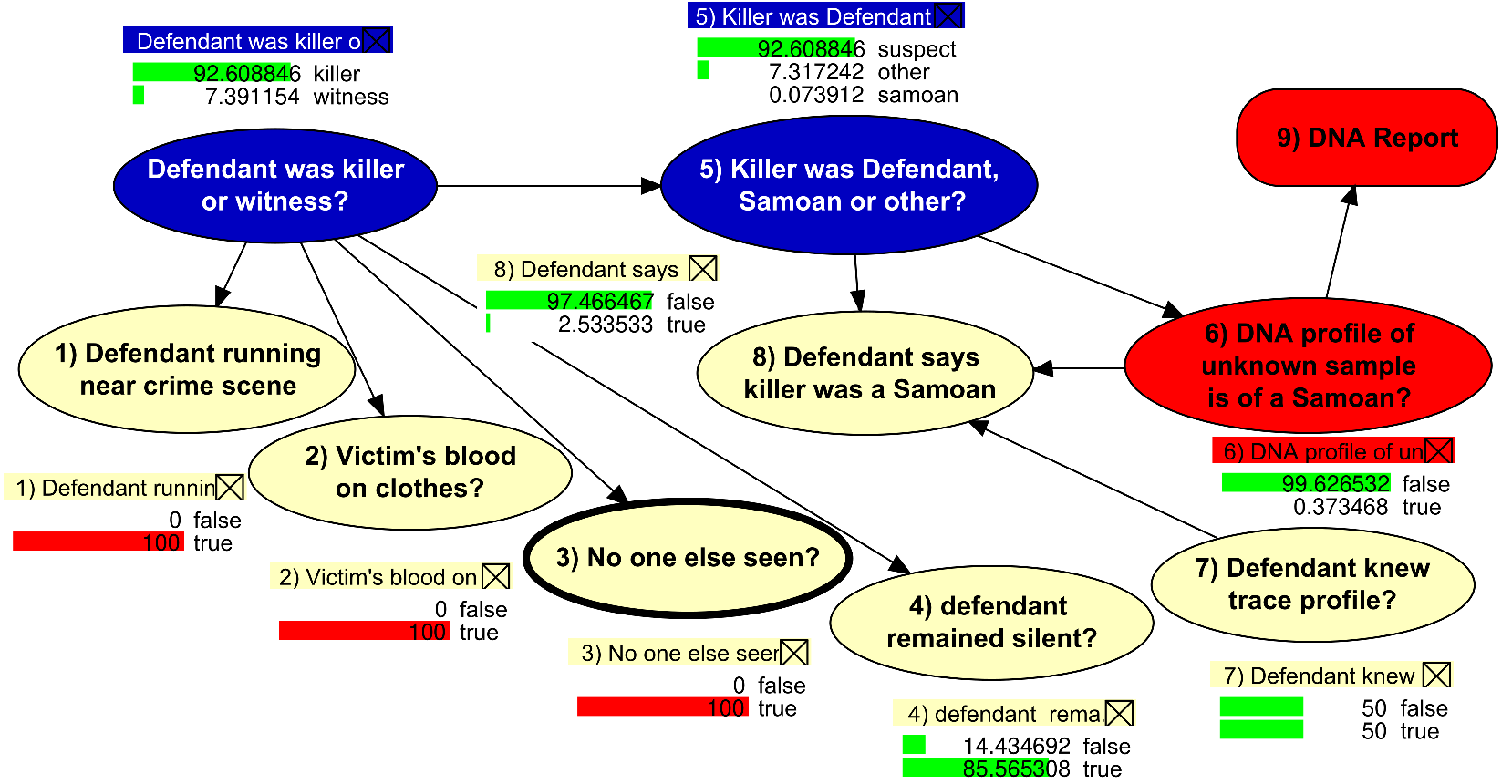}
	\end{center}
\end{figure}

\begin{figure}[htbp]
	\begin{center}
		\caption{Updated probabilities conditioned on the additional  evidence that the defendant remains silents.} 
		\label{fig:silent}
		\includegraphics[width=.90\textwidth]{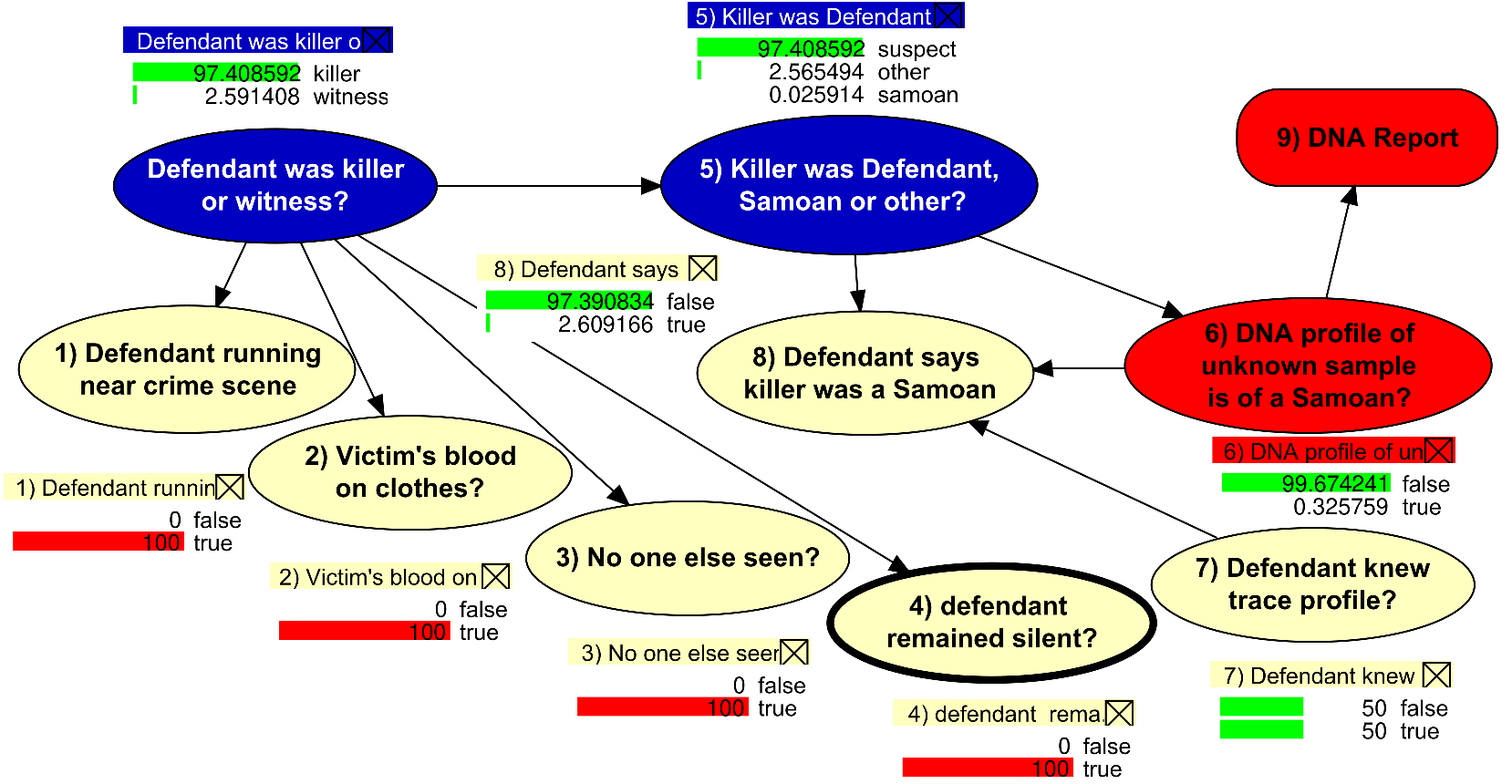}
	\end{center}
\end{figure}

\begin{figure}[htbp]
   	\begin{center}
   		\caption{Updated probabilities conditioned on previous evidence and that the defendant says a Samoan was the murderer.} 
   		\label{fig:Samoan}
   		\includegraphics[width=.90\textwidth]{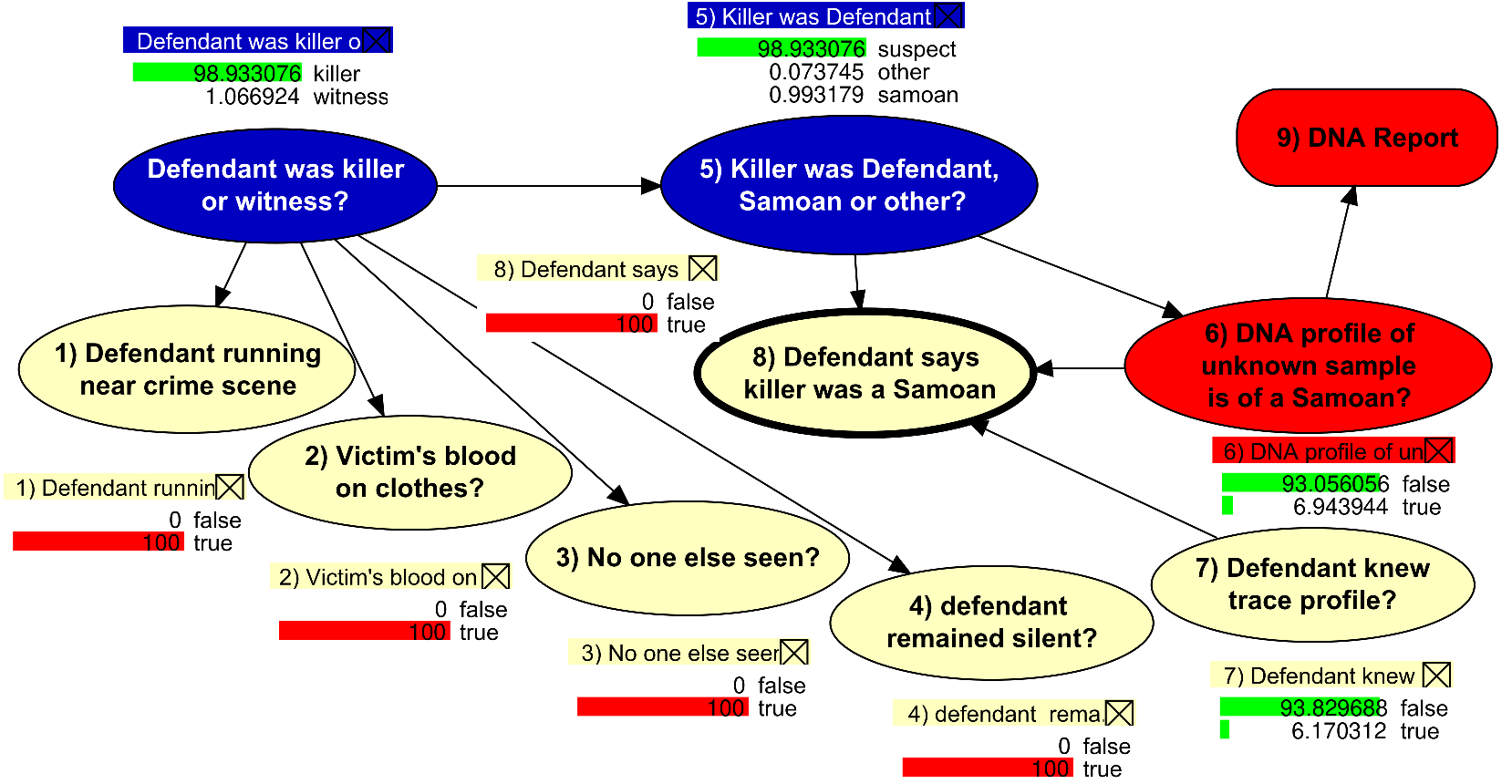}
   	\end{center}
   \end{figure}

\begin{figure}[htbp]
   	\begin{center}
   		\caption{Updated probabilities conditioned on previous evidence and the results of the DNA analysis.} 
   		\label{fig:SamDNA}
   		\includegraphics[width=.90\textwidth]{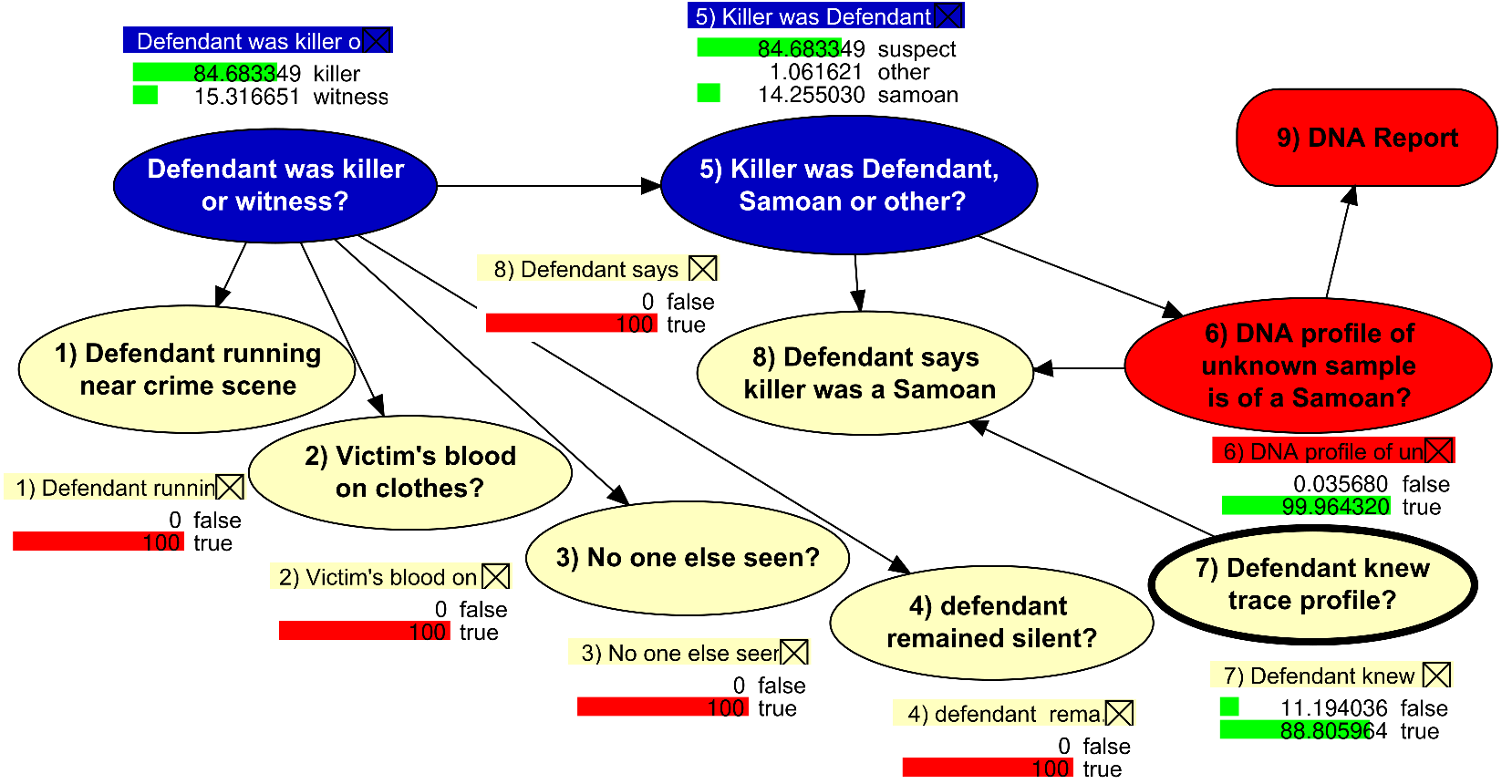}
   	\end{center}
   \end{figure}

\begin{figure}[htbp]
	\begin{center}
		\caption{Bayesian network after propagating all the evidence when defendant does not know about the DNA report.}
		\label{fig:notknow}
		\includegraphics[width=.90\textwidth]{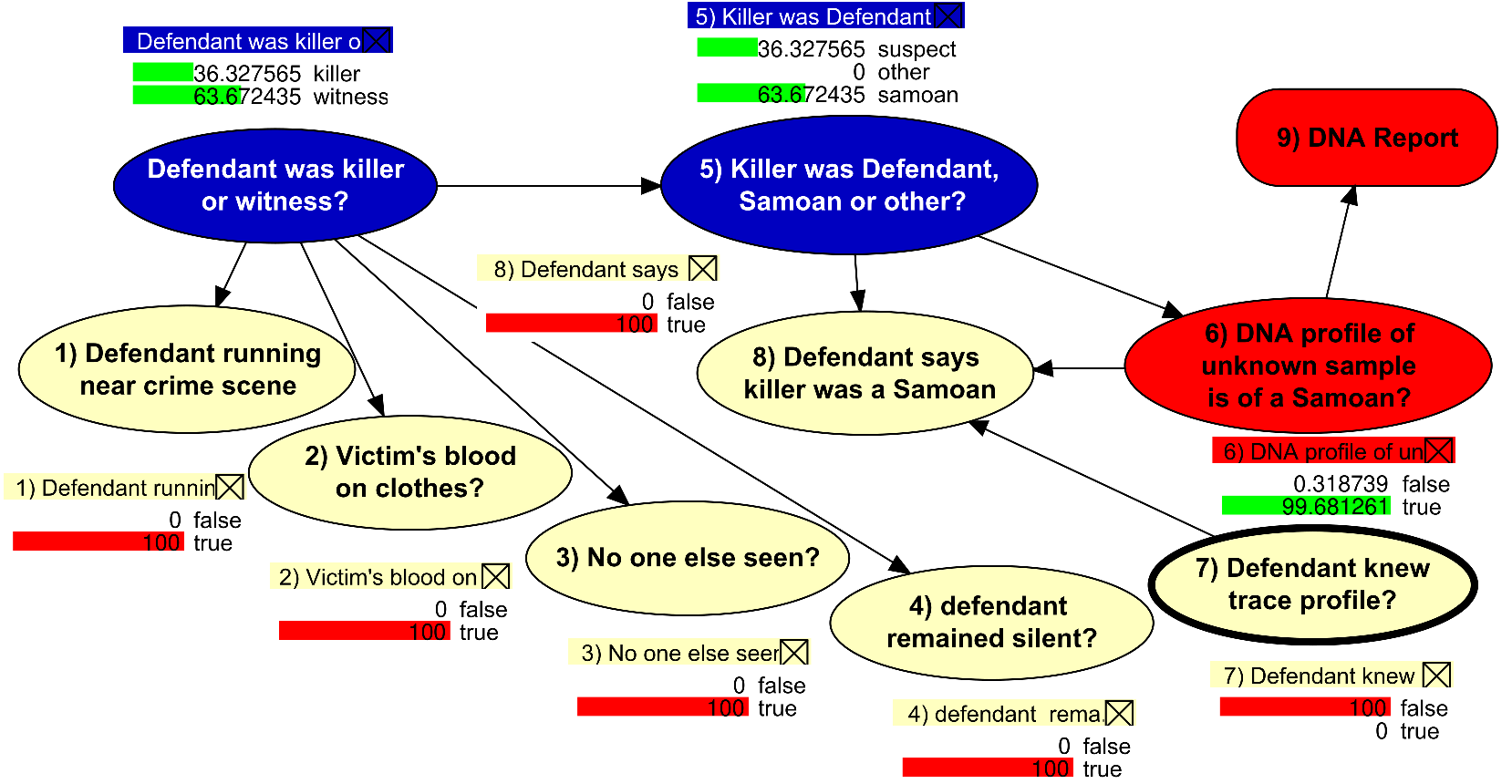}
	\end{center}
\end{figure}

\begin{figure}[htbp]
		\begin{center}
			\caption{Bayesian network  after propagating the  DNA  and  testimony evidence when defendant  knows beforehand about the DNA report.} 
			\label{fig:knows}
			\includegraphics[width=.90\textwidth]{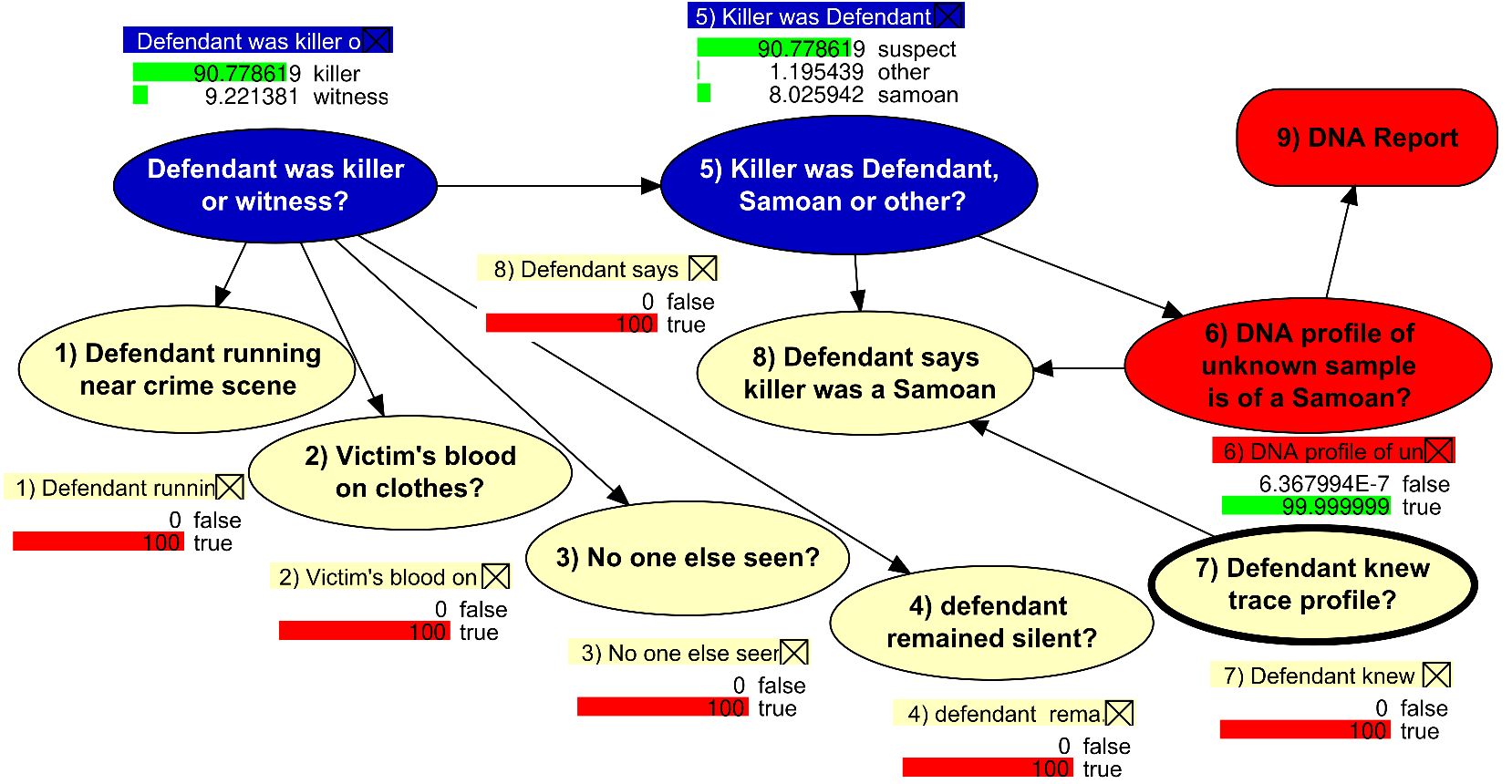}
		\end{center}
	\end{figure}

Once the conditional probabilities are assigned, it is possible use the Bayesian network as a tool to evaluate the strength of the observed evidence against the defendant.  We illustrate this process in Figures \ref{fig:naive}--\ref{fig:knows}.  
Figure \ref{fig:naive} shows the initial state of the BN, before any of the variables are instantiated.  It shows the marginal probabilities that the model assigns to all states of each node,  before considering what the evidence shows.   The top-left  node has no parents, and we assign equal prior probabilities that the defendant is the killer or merely a witness of the crime.   

 The marginal probabilities for any child node are calculated  based on the marginal probabilities of the parent nodes and the conditional probabilities of the node in question. For example, before we know whether the victim's blood was found on the defendant's clothing, the BN assigns a marginal probability of that event at 85\%.  This number should make sense intuitively because we put the conditional probability of finding the victim's blood at 0.9 if defendant was the killer and 0.8 if he was a witness, and we deemed these two events equally probable (at this initial stage).

Figures \ref{fig:run}--\ref{fig:noone} show how the hypotheses on whom is the culprit change as we instantiate each of these items of evidence to their true state.  The probability the defendant was the killer (shown in the upper left node) increases roughly to 55\% based on \textbf{1)}  the defendant was running near the scene (Figure \ref{fig:run}); then to 58\% if we also consider \textbf{2)} finding the victim's blood on his clothing (Figure \ref{fig:blood}), then to 93\% when we also consider \textbf{3)} that no one else was seen in the area (Figure \ref{fig:noone}), and to 97\% when we consider \textbf{4)} his decision to remain silent (Figure \ref{fig:silent}).  The behavior of the model is straightforward and should be easy to understand.    These four items of evidence are conditionally independent prior to evidence but marginally dependent once evidence is instantiated in any one of the 4 items.  

Each item is incriminating because it is more probable under the theory that defendant is the killer than the theory that he was a witness.  So, each adds to the strength of the overall case against the defendant, increasing the posterior probability that he is guilty.   
Figure \ref{fig:Samoan} shows what happens when we also consider the defendant's claim that the killer was a Samoan.  This claim, by itself, causes an increase in the posterior probability that the killer was the defendant (to 99\%), as shown in node \textbf{5)}  in Figure \ref{fig:Samoan}. However, when we also propagate the evidence on the DNA this then causes it to decrease to 84\%, as shown in Figure \ref{fig:SamDNA}.
   
 The probability of guilt drops to 36\% (Figure \ref{fig:notknow}), if we assume that the defendant did not know about the DNA evidence before claiming the killer was Samoan.  The DNA test and the defendant's statement both independently support the same theory --that the killer was a Samoan-- which makes it look more likely that this hypothesis is true.  This dramatic reduction in the probability of guilt depends critically, however, on the assumption that the defendant was uninformed about the DNA test.  If instead the defendant knew about the DNA test, then (as shown in Figure \ref{fig:knows}) the probability of guilt is  over 90\%.

\section{Summary of the analysis of the DNA evidence}
\label{sec:DNA}

Here we give a brief summary of the analysis of the DNA evidence. A more detailed account is given in Appendix 2 and 2a. 

In the case we analyse here  there is uncertainty about which population is
relevant for the analysis of the sample from the unknown actor. In general, the unobserved actor/actors are assumed to have
genes drawn from a specific reference  population,  and the results can depend on which
population (and corresponding allele frequency database) is used.
When  there is uncertainty about which  subpopulation  
unknown actors belong to, dependence between genetic markers is induced.  
This issue is most often ignored in forensic DNA analysis.
Additionally, uncertainty  about the relevant population,  can induce dependence between actors, observed or not \citep{green:mortera:09}.

In this crime case the potential individual whose DNA was found at the crime scene could either be  of  Samoan origin or  alternatively of Hispanic, Caucasian or Afro-American origin, each having different allele frequencies.   In  \citet{green:mortera:09} we show that mixing across subpopulations is not the same as averaging the allele frequencies and assuming an undivided subpopulation. This observation may have wider implications; since
all real populations are to a degree heterogeneous and in this way 
dependence between markers will always be present.

Table \ref{tab:LR} in Appendix 2a shows that the exact likelihood ratio given by the BN is roughly 302, whereas, if we  calculate  the overall likelihood ratio  as a  product of  the likelihood ratios for each marker, assuming incorrectly that the markers are independent,  we get a much smaller value. This illustrates that,   considering potential different subpopulations or heterogeneity, induces dependence between markers. This fact is very often overlooked in forensic DNA analysis and can lead to either more or less incriminating results.

%
\section{Discussion and Concluding Remarks}
\label{sec:conc}

Let’s now return to the philosophical question that motivated this exercise:  Does the epistemic support for a theory rest solely on the logical relationship between the theory and the evidence?  Or, does it depend partly on how the theory was developed and tested?  As noted earlier, this question has perplexed and divided philosophers for a very long time.  Our case study casts fresh light on this question.  

Our Bayesian analysis assumes that the epistemic support for a theory rests solely on the logical relationship between the theory and the evidence.  It makes the further assumption that the strength of the evidence for supporting the theory—it’s probative value—depends solely on the relative probability of the evidence under the theories being evaluated.  Based on these assumptions, supporters of the “logical theory of confirmation” (e.g., Hempel, 1945; Glymour, 1980; Edwards, 1972; Royall, 1997) have argued that the strength of epistemic support provided by evidence is unaffected by whether it was collected before or after the theory was developed.  We disagree because, as we demonstrated, the order in which the evidence is discovered, relative to theory development, can, in some instances, change the conditional probability of the evidence under the theories being evaluated and thereby alter its probative value.  Hence, the same item of evidence may be valued differently if it is “old evidence” that was accommodated in theory development than if it was “new evidence” collected later, although the distinction between “old” and “new” evidence is not always important.  

Consider, for example, the evidence that defendant was running near the scene of the crime shortly after the murder.  We believe the conditional probability of this item is a bit higher if the defendant was the killer than if he were merely a witness.  We do not, however, believe that either conditional probability depends on how or when the theories of the case were developed.  Whether this item is “old evidence” that the defendant knew about before offering his theory, or “new evidence” that was discovered later, makes no difference to the conditional probability of the evidence under the prosecution or defence theories.  

We believe the same is true of most of other evidence items, including the blood matching the victim’s DNA profile on defendant’s clothing, that no one else was seen in the area, and that defendant chose to remain silent when questioned.  The probability of each of these items obviously depends on whether the prosecution or defence hypothesis of the case is true; but it does not depend on when the prosecution or defence theories were developed relative to the discovery of this evidence.  Whether these items are “old evidence” or “new evidence” does not affect the strength of support that they provide for the prosecution or defence case.

The sequence in which the evidence and theory emerged matters tremendously, however, when we consider the DNA report and the defendant’s claim that the killer was a Samoan.  Let’s consider two possible sequences in which the evidence and theory may have developed: 

\textbf{Sequence One}--the defendant first reported that the killer was a Samoan, this report led the defence lawyer to present the theory that the killer was a Samoan, and the DNA evidence later supported the Samoan theory;

\textbf{Sequence Two}--the DNA report appeared first, and the defendant claimed the killer was a Samoan only after learning that the DNA report supported that theory.

In Sequence Two, defendant’s theory may have been shaped to fit the DNA evidence; and the evidence (defendant’s statement that the killer was a Samoan) may have been shaped to fit the theory.  Neither of these processes could have occurred in Sequence One.  

The Bayesian analysis shows why the epistemic support for the defence case is stronger in Sequence One than in Sequence Two.  The difference arises entirely from the way in which the sequence affects the conditional probability of the defendant’s statement under the prosecution and defence theories of the case.  In Sequence One, there is a far greater chance that the defendant would claim the killer was a Samoan if the killer was actually a Samoan than if the killer was someone else (including himself).  Consequently, his statement is  highly probative.  Moreover, it is conditionally independent of the other evidence supporting his innocence, which means that it adds additional value beyond what the other evidence shows.  In Sequence Two, the defendant has reason to lie about the killer being a Samoan, hence the probability he would make such a statement is not much different if he is lying than if he is telling the truth. In Bayesian terms, the defendant’s statement is conditionally dependent on the DNA report, and its incremental value--the value it adds beyond the other evidence—is minimal.  

The distinction between “old” and “new” evidence is not particularly helpful in explaining how the sequence in which the evidence and theory emerged affects the strength of the case.  In Sequence One the defendant’s statement is “old evidence” and the DNA report is “new evidence;” in Sequence Two, the DNA report is “old evidence” and the defendant’s statement is “new evidence.”   So, our analysis does not support claims that “new evidence” provides stronger epistemic support than “old evidence,” or vice-versa.  Instead it shows that the epistemic support provided by a body of evidence depends on the inferential connections among the items of evidence and between the evidence and the theories being evaluated.  Whether the theory or the evidence came first sometimes affects those inferential connections and sometimes does not.  

When might we expect the order of theory development, relative to the discovery of evidence, to affect the strength of support for a theory?  And when might we expect it to have no effect on strength of support?   The question can be approached in a case-by-case manner by using Bayesian networks, as we have done in our case study.  Construction of a Bayesian network requires one to think systematically about the conditional probability of the evidence under the theories and about various factors that may affect those conditional probabilities.  It may well reveal situations in which the conditional probabilities vary depending on who knew what, and when they knew it.  Those are the situations in which the strength of support for a theory may depend on how and when the theory was developed, relative to when the evidence became known. 

While each case will require a separate analysis, we anticipate that the order of theory development vis-a-vis evidence collection will be most important in two general circumstances: (1) where the collection or reporting of evidence may have been shaped to fit the theory; and (2) where the theory may have been shaped in inappropriate ways to fit the evidence.  

\subsection{Shaping Evidence to Fit Theory}  

Observer effects have long been noted in science \citep{risinger2002daubert}.  What is observed and recorded by witnesses in criminal cases, and by scientific researchers, may be affected by expectations and desires of the observer—and these biasing effects can occur without the observer intending or even being aware of it.  Consequently, when evaluating the strength of the body of evidence supporting a theory, it is important to consider whether expectation or desires, arising from investigators’ knowledge of the theory, or of evidence supporting the theory, may have created bias in the collection or reporting of new evidence.  These effects, sometimes labeled contextual bias, can undermine the value of reported observations by changing the conditional probability of those reports under the relevant theories—typically causing the reports to be slanted in a manner that supports theory-based expectations \citep{cooper2019cognitive,kassin2013forensic}.  

\citet{thompson2023shifting} used a Bayesian network analysis to show how contextual bias can weaken the probative value of evidence against a criminal defendant.  In a simple case that depended on two pieces of evidence—an eyewitness who identified the defendant, and a forensic scientist who linked the defendant to the crime scene with pattern-matching evidence—the case was strongest when the two evidence items were conditionally independent of one another.  The strength of the case was weaker when there were conditional dependencies between the evidence items, which might arise if, for example, the forensic scientist became more likely to incriminate the defendant if he knew that the defendant had been identified by the eyewitness, or vice-versa.  

Contextual bias can also arise in scientific research.  Historians of science claim to have found examples of data being slanted, perhaps inadvertently, toward theory-confirmation even in the work of such prominent figures as Isaac Newton and Gregor Mendel.  While such problems can often be mitigated by using blinding and “unmasking” procedures that shield those collecting and interpreting the data from potentially biasing information, those evaluating the strength of scientific evidence must always consider whether such procedures were needed and, if so, were used effectively.  When such a bias is possible, it may well affect the conditional probability of the evidence under the relevant theories in ways that reduce the epistemic strength of the evidence.  In these cases an item of evidence may be weaker if it is “new” (collected post-theory, and perhaps shaped to fit, the theory), than if it was “old” (collected before the theory emerged).

Bias in the collection and interpretation of data is typically the product of poor scientific practices rather than an intent to mislead or deceive.  Intentional fraud is not unknown, however, and is another way in which evidence may be shaped to fit the expectations or desires of investigators \citep{hand2007deception}. A criminal defendant has an obvious motive to fabricate support for helpful theories of a case (if and when he knows in advance what those theories are).  The same problem can arise in scientific research if individuals reporting relevant findings have a professional, reputational or commercial stake in supporting a particular theory and know in advance what findings will be helpful.  In these instances, knowledge of “old evidence” may induce individual to misrepresent facts in a manner designed to support a particular theory may undermine the veracity of “new evidence” in ways that reduce the overall strength of support for the theory.

\subsection{Shaping Theory to Fit Evidence. }

Scientists and criminal investigators always try to shape their theories to accommodate existing evidence—that process is essential to theory development.  In some instances, however, the way in which a theory is developed may affect the strength of epistemic support for the theory.  Philosophers have argued that questionable practices in theory development, such as overfitting, hypotheses hunting, and “fudging” of a theory to fit a particular datum, should reduce confidence in a theory  \citep{syrjanen2023novel}.  We agree and we believe this effect can be captured in Bayesian network models. 

Overfitting occurs when researchers “fit” a statistical model to data from a limited sample, and then try to apply that model to the broader population from which the sample was drawn.  Rarely does the model fit the entire population, or other samples drawn from that population, as well as it fits the original sample on which it is based.  The problem is that such models incorporate features of the sample data that arise by chance and hence do not accurately reflect the characteristics of the underlying population.  

Suppose, for example, that researchers conducting a longitudinal health study on a limited sample of participants find a strong relationship between consumption of kumquats and digestive health.  To the extent this finding may have arisen from  spurious correlation, it should be viewed with skepticism.  This is particularly true if the researchers discovered the relationship between kumquats and digestive health by “hypothesis hunting”—that is, by scanning through their data looking for interesting findings.  It is also true if the researchers “fudged” or changed their theory to fit the data—for example, if they planned to test the hypothesis that consumption of fruit improves digestive health, found limited support, and then modified the theory to focus on kumquats so they would have statistically significant findings to report.  Researchers who develop theory in this manner are effectively painting the target around the arrow, which increases the probability that the findings supporting the theory are products of spurious relationships and products of data-dredging  \citep{thompson2009painting}.  In Bayesian terms, the conditional probability of these findings under the kumquats-are-helpful theory may be lower in reality than suggested by the data in the initial sample.  These concerns would be greatly reduced, however, if the theory that kumquats improve digestive health was developed on independent grounds and the study was conducted later in order to test that theory.  Hence, these are instances in which “new” data may be more probative than “old” data.

\subsection{Conclusions}

When we began this project, we planned to focus on theory confirmation in criminal justice, drawing support and advice from philosophy.  Following the lead of Mackor, we hoped that the insights of epistemologists and philosophers of science would guide us in developing better Bayesian models.  While we have certainly benefited from that body of work, we have come to believe that our Bayesian network approach might also be helpful to philosophy.  Some of the conundrums that have made it difficult for philosophers to agree on the relative merits of prediction and accommodation, such as the “Bayesian problem of old evidence” \citep{glymour1980theory} become easier to understand and address when modeled using Bayesian networks (see Appendix 1).  A Bayesian network can be viewed as an argument about the strength of evidence for evaluating theories or hypotheses.  It is an argument that requires one to be rigorous and precise about the nature of the connections among items of evidence and between the evidence and hypotheses.  In the process of developing our Bayesian model it has become clearer that no general conclusions can be drawn about the epistemic strength of evidence for supporting a theory based on whether the evidence is “new” or “old.”   What is needed instead is careful consideration of how the theories in question may have shaped the evidence offered to support them, and vice-versa—as well as how that shaping process affected the conditional probability of the evidence under those theories. We have shown how that approach is helpful for analyzing the strength of evidence in a criminal case; we hope it also proves helpful to those engaged in broader philosophical discussions of theory confirmation.

\section*{Appendix 1: : Is there a “Bayesian problem of old evidence”?}
\label{sec:appendix 1}

Some philosophers believe that Bayesian analysis does not allow old evidence to confirm new theory.  In a widely-cited chapter titled “Why I am Not a Bayesian,” Clark Glymour explained the matter as follows:    

\begin{quote}
	Scientists commonly argue for their theories from evidence known long before the theories were introduced.  Copernicus argued for his theory using observations made over millennia$\ldots \ldots$Newton argued for universal gravitation using Kepler’s second and third laws, established before the Principia was published.  The argument that Einstein gave in 1915 for his gravitational field equations was that they explained the anomalous advance of the perihelion of Mercury, established more than half a century earlier$\ldots \ldots$. Old evidence can in fact confirm new theory, but according to Bayesian kinematics, it cannot. \citep{glymour1980theory} 
\end{quote}
    Glymour relied on a mathematical argument to support this claim.  Let $T$ represent a theory such as Einstein's general theory of relativity.  Let $E$ represent an item of evidence supporting that theory, such as the anomalous advance of the perihelion of Mercury.  Bayesian logic requires that $p(T|E) = P(T) \times p(E|T)/p(E)$.  If $E$ is “old evidence” that is known to be true, however, then according to Glymour, $p(E)$ and $p(E|T)$ will both equal one, which means that $p(T|E) = p(T)$, and consequently that $E$ has no probative value.  This “Bayesian” conclusion is obviously wrong, of course, because scientists did indeed treat the advance of Mercury’s perihelion as supportive of Einstein's theory.

       Glymour’s argument came to be known as the   “Bayesian problem of old evidence” \citep{barnes2022prediction}.  It spawned a large literature  \citep{christensen1999measuring, Eells2000,barnes1999quantitative,hartmann2015new}, with some commentators suggesting that the “problem” can only be solved via an elaboration on the Bayesian approach (e.g. \citet{howson1991old}).  

  From our perspective, there was never an actual problem with Bayesian analysis.  Glymour's perception of a problem arises from a misunderstanding of two key aspects of how Bayesian analysis works. First, Glymour focused on a single hypothesis, while Bayesian analysis is typically used to distinguish between two (or more) alternative hypotheses.  Second, Glymour failed to recognize the subjunctive nature of the conditional probabilities used in Bayesian analysis.  When evaluating $p(E|T)$, Bayesians are asking “how likely would $E$ be if $T$ were true.”  This allows Bayesians to assign a value far less than one to $p(E)$ and to $p(E|T)$ even if they know $E$ has occurred.  For example, suppose that $E$ is evidence that John won a lottery, and $T$ is the theory that the lottery was fair and random.  The fact that we know he won does not prevent us from assigning a low value to $p(E|T)$, if we think it is unlikely that he would have won if the lottery was fair.  Furthermore, because the law of total probability requires that $p(E) = p(E|T)p(T) + p(E|\overline{T})p(\overline{T})$, where $\overline{T}$ stands for the proposition that $T$ is not true, the value we assign to $p(E)$ should also be less than one so long as we believe $p(E|T)$ is low.

  Alternatively, suppose we use  use a screening test for  a disease $D$ (say  1\% of the population has the disease) and  a blood sample tests positive $E$. Even if we observe a positive test, this  does not imply that $p(E)=p(E|D)$ are both certain. As customary for any screening test, suppose the test  has the following features:
  \begin{itemize}
  	\item[(i)] the probability of correctly detecting the disease, if present $p(E|D)=0.99$ (Sensitivity of the test)
  	\item[(ii)] the probability of correctly not detecting a disease, if
  	absent $p(\overline{E}|\overline{D})=0.95$. (Specificity of the test)
    \end{itemize}
  Then by the law of total probability $p(E) = p(E|D)p(D) + p(E|\overline{D})p(\overline{D})=0.06$ is far less than 1. Furthermore,  by Bayes' theorem  the probability that the person whose blood was tested positive has the disease is roughly $p(D|E)= 0.17$ certainly less than 1?

In this light, let's return to the role of “old evidence” in supporting Einstein’s general theory of relativity ($GTR$).  A proper Bayesian analysis would consider the conditional probability of $E$ (the advance of Mercury’s perihelion) under the competing theories scientists were trying to distinguish--$GTR$ and Newtonian gravitational theory ($NGT$).  Under $GTR$, the probability of $E$ was very close to one.  Under $NGT$, however, $E$ is difficult to explain and hence, stated in the subjunctive, $E$ would be improbable were $NGT$ true.  Consequently, a Bayesian analysis suggests that $p(E|GTR)$ is much higher than $p(E|NGT)$, and consequently that $E$ provides strong support for $GTR$ over $NGT$.   

Scholars have suggested that Einstein took account of $E$ when coming up with $GTR$ (\citet{gong2022learning},\citet{earman1978einstein}, p. 300), and hence derived the theory in part from the very data it purported to explain.  As discussed in the text, this process may risk overfitting a theory to “old evidence” which might cause one to overestimate the probability of such evidence under the theory.  Even so, scholars of the time, 1915, would have assigned far higher values to $p(E|GTR)$ than to $p(E|NGT)$.  In other words, a proper Bayesian analysis shows that this “old evidence” has considerable probative value for establishing $GTR$, just as scientists of the time assumed that it did.  Any concerns that might have been raised about “theory fudging” in connection with $GTR$ have, of course, been answered by subsequent confirmatory findings-- e.g., on the curvature of light in gravitational fields.  
 \citet{jefferys1992ockham} showed that the Bayes factor 
$p(E|GTR)/p(E|N)$  gives the odds in favour of Einstein's theory compared to $NGT$ of about 30:1.

\section*{Appendix 2: Analysis of the DNA profile}
\label{sec:appendix 2}

The evidence $E$ found at the scene of the crime  consists of  the DNA profile given 
in \tabref{profile}. The corresponding markers, alleles and  allele frequencies of the crime profile in  different populations is given in 
\tabref{freq}.
\begin{table}[htbp]
	\centering
	\caption{DNA profile at scene of the crime}
	\begin{tabular}{l|rr|rr|rr|rr|rr|rr|}
		marker  & \multicolumn{2}{c|}{D2}	& \multicolumn{2}{c|}{CSF}	 & \multicolumn{2}{c|}{D7} 	& \multicolumn{2}{c|}{D21}	&	\multicolumn{2}{c|}{D8} & \multicolumn{2}{c|}{D16} \\
		\hline
		alleles & 18 & 22  & 11 & 14 & 12 & 12 & 28 & 34.2 & 10 & 10 & 14 & 14 \\
	\end{tabular}%
	\label{tab:profile}%
\end{table}%

\begin{table}[htbp]
	\centering
	\caption{Markers, alleles and corresponding allele frequencies of the crime profile in the different populations.}
	\begin{tabular}{lr|rrrr}
		marker & \multicolumn{1}{l}{allele} & \multicolumn{1}{l}{Samoan} & \multicolumn{1}{l}{Hispanic} & \multicolumn{1}{l}{Caucasian} & \multicolumn{1}{l}{AfroAmerican} \\
		\hline
		D2    & 18    & 0.12 & 0.08 & 0.073 & 0.04 \\
		& 22    & 0.25 & 0.057 & 0.034 & 0.14 \\
		CSF   & 11    & 0.39  & 0.28  & 0.31  & 0.25 \\
		& 14    & 0.01  & 0.006 & 0.01 & 0.009 \\
		D7    & 12    & 0.22  & 0.15  & 0.16  & 0.088 \\
		D21   & 28    & 0.26  & 0.10   & 0.16  & 0.25 \\
		& 34.2  & 0.016 & 0.005 & 0.004 & 0.003 \\
		D8    & 10    & 0.21  & 0.093 & 0.1   & 0.03 \\
		D16   & 14    & 0.12 & 0.13 & 0.026 & 0.025 \\
	\end{tabular}%
	\label{tab:freq}%
\end{table}%

Table 6 gives the racial and ethnic composition in  Long Beach, California
based on the US Census in 2000 around the time when  the crime  was committed.
Note that there were roughly 1\% Pacific Islanders.
\begin{table}[htbp]
		\centering
	\caption{Racial and ethnic composition in  Long Beach, California in 2000. NH stands for Non Hispanic}
	\begin{tabular}{l|rr}
		\hline
		{Race / Ethnicity} & {Population} & {Percentage} \\
		\hline
		White  (NH) &    152,899 &    33.13\% \\
		Black or African American  (NH) &     66,836 &    14.48\% \\
		Native American or Alaska Native  (NH) &      1,772 &     0.38\% \\
		Asian  (NH) &     54,937 &    11.90\% \\
		Pacific Islander  (NH) &      5,392 &     1.17\% \\
		Other race  (NH) &      1,013 &     0.22\% \\
		Mixed race or Multiracial (NH) &     13,581 &     2.94\% \\
		Hispanic or Latino (any race) &    165,092 &    35.77\% \\
		\hline
	\end{tabular} 
	\label{tab:Census} 
\end{table}

%

\begin{figure}[htbp]
	\begin{center}
		\caption{Bayesian network representing the class network \textbf{DNA report}}
		\label{fig:master}
		\resizebox{0.9\textwidth}{!}{\includegraphics{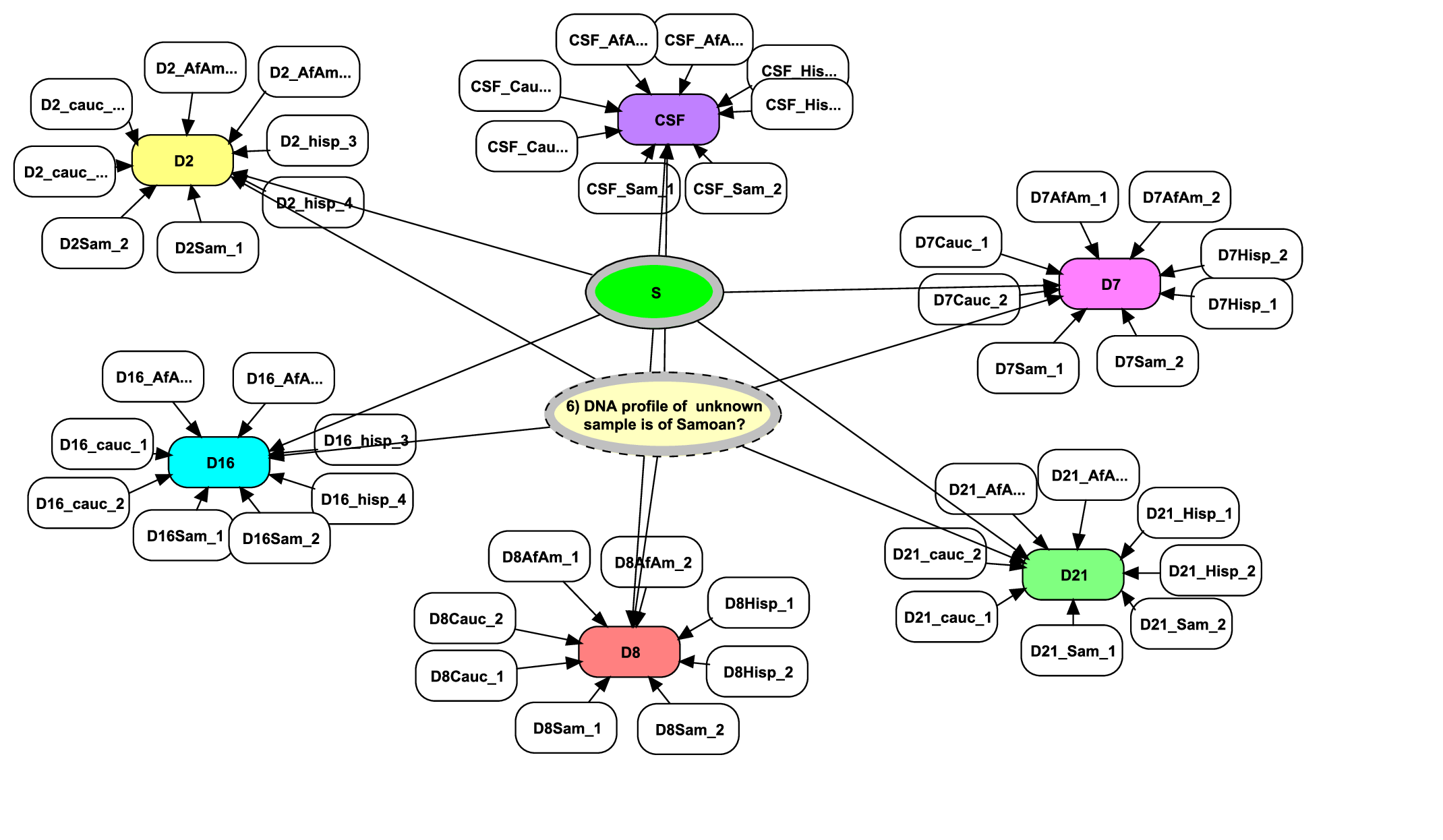}}
	\end{center}
\end{figure}

\figref{global} in Section \ref{sec:BN} represents the ``top-level'' Object-Oriented Bayesian Network  (OOBN), constructed using the
graphical interface to \Hugin.  All nodes already described  are simple nodes -- except the node representing \textbf{DNA Report}  which has further
internal structure we will describe here. 
\figref{master} represents the underlying structure of the network   \textbf{DNA report}, where in turn each marker network,  \textbf{D2}, \textbf{CSF}, \textbf{D7}, \etc\  is an instance of a network as in \figref{marker}, which is explained in Appendix 2a.

\begin{figure}[htbp]
	\begin{center}
		\caption{Network for one marker.} 
		\label{fig:marker}
		\includegraphics[width=1.0\textwidth]{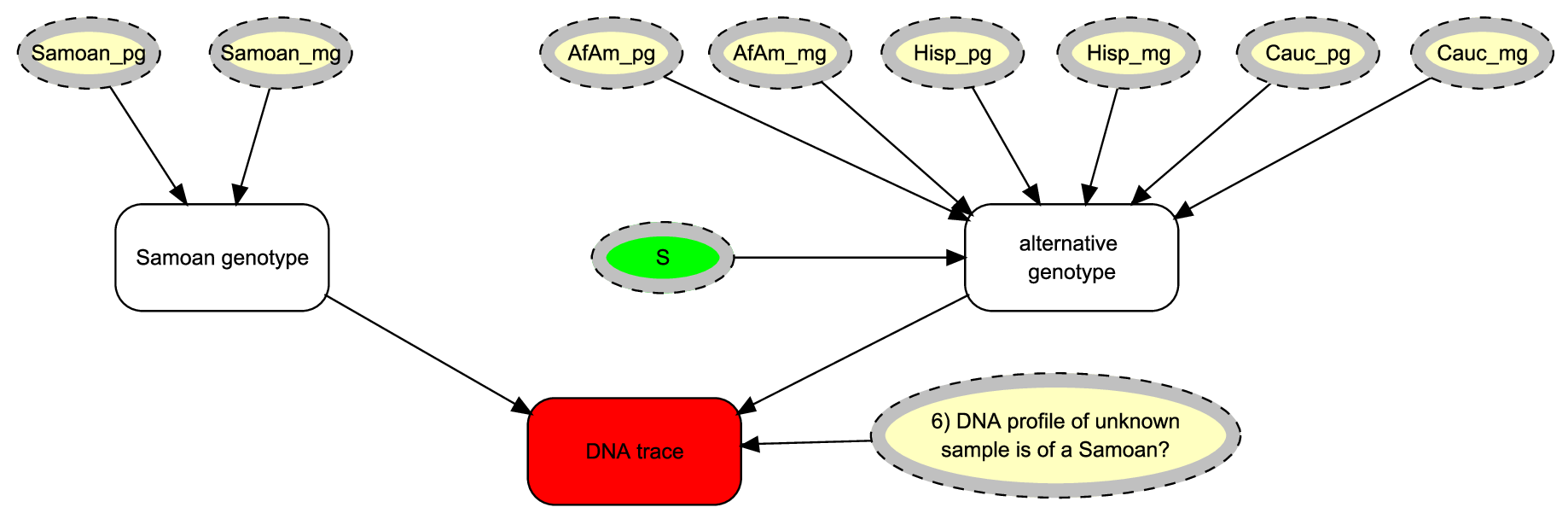}
	\end{center}
\end{figure}

\subsection*{Appendix 2a: Population heterogeneity (HET)}
\label{sec:HET}

Uncertainty  about the relevant population,  can induce dependence between actors, observed or not. To account for this uncertainty, as in \citet{green:mortera:09}
 we can construct the network for each marker as in \figref{marker}.

We consider  both the Samoan subpopulation, and potential other reference  populations, \ie\ 
Afro-American, Hispanic and Caucasian subpopulations. 
Node 6) \texttt{DNA profile of unknown sample is of a Samoan?} selects, according to its \textit{true/false} state, between the two alternative explanations: a) an unidentified  Samoan left the trace at the crime scene;  or b)  an alternative person of unknown ethnicity left the trace.
Here we assume this alternative person  could be from 
Afro-American, Hispanic or Caucasian subpopulations. The green node   {\tt S} takes values $0,1,2$ corresponding to the Hispanic, Caucasian and  Afro-American  populations having prior probabilities proportional to the ethnic composition as in Table 6.  The  allele frequencies for each population  correspond to those in Long Beach as in Table 5. These frequencies populate the top nodes in Figure \ref{fig:marker} which represent the maternal and paternal  genes.


For further details on the structuring of the networks refer to the detailed documentation in \begin{verbatim}https://www.dropbox.com/scl/fo/zvxth8gnfnzxbep7z7kyt/AOh-AHx-q6v5rSi2kVpe3jY?rlkey
	=t9q9ndsjr7nebpp3nlltjc8po&dl=0
\end{verbatim}

The node {\tt S} in Figure \ref{fig:marker} is a variable identifying the
subpopulation, which is dependent  between actors.
As in  \citet{green:mortera:09}, for each actor, {\tt S} is the same for both genes for
all markers, so that mixing across subpopulations is not the same
as averaging the allele frequencies and assuming an undivided
subpopulation. 

\begin{figure}[htbp]
	\begin{center}
		\caption{Overall Network when propagating only the DNA evidence.} 
		\label{fig:OnlyDNAev}
		\includegraphics[width=1.0\textwidth]{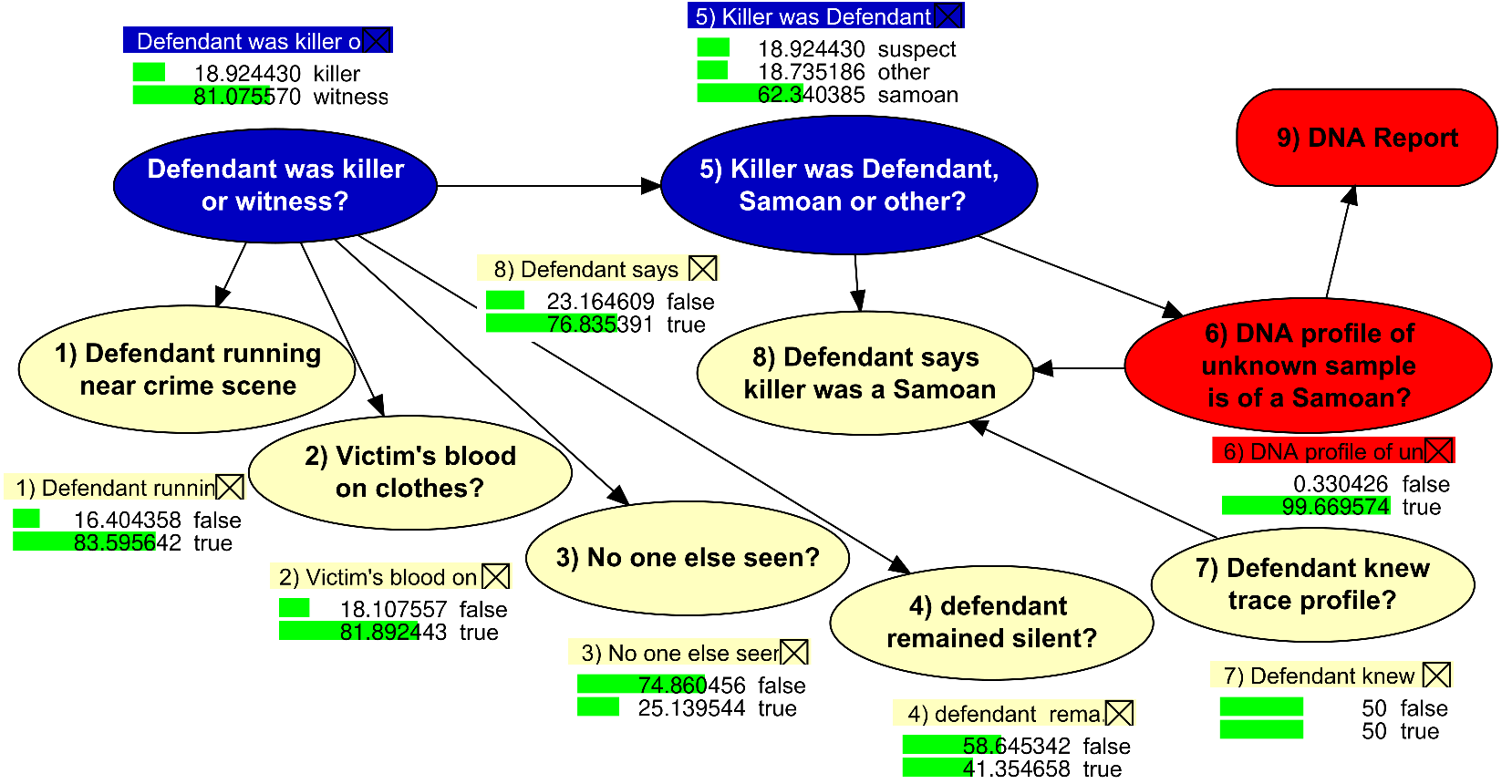}
	\end{center}
\end{figure}

\begin{table}[htbp]
	\centering
	\caption{Likelihood ratio for each marker, the  exact \LR\ and the \LR\ using the product rule.}
	\begin{tabular}{|l|rrr|}
		\hline
		& \multicolumn{1}{l}{$\Pr(E \cd \Hp )$} & \multicolumn{1}{l}{$\Pr(E \cd \H_d)$ } 
		& \multicolumn{1}{c|}{{ $\LR_i$ }} \\
		\hline
		D2     &      0.573 &      0.427 &      1.34 \\
		CSF &      0.662 &      0.338 &      1.96 \\
		D7 &      0.299 &      0.701 &      0.43 \\
		D21 &      0.882 &      0.118 &      7.49 \\
		D8 &      0.536 &      0.464 &      1.15 \\
		D16 &      0.952 &      0.048 &     19.92 \\
		\hline
		product rule &            &            &   163.02 \\
		exact &       0.997 &      0.0033 &    301.6 \\
		\hline
	\end{tabular}%
	\label{tab:LR}%
\end{table}%


When  the evidence from all the markers is propagated in the global network as in \figref{OnlyDNAev},  the posterior probability of node 6) that the DNA is of Samoan origin is  0.9967. This gives  the exact likelihood ratio (LR) of 302.  
Table \ref{tab:LR} shows that, if instead of computing the exact likelihood ratio,  we  calculate  the overall likelihood ratio  as a  product of  the likelihood ratios for each marker, $\LR=\prod_i \LR_i = 163$, we obtain a much smaller  and incorrect value. This illustrates that  when considering potential different subpopulations or heterogeneity of a reference population, this induces dependence between markers.


%
%

\bibliographystyle{apalike}
\bibliography{Samoan}
\end{document}